\documentclass{webofc}

\usepackage[varg]{txfonts}   
\usepackage{hyperref}
\usepackage{url}
\def \mykzeros{$K_{S}^{0}$\xspace}
\def \mydedx{d\textit{E}/d\textit{x}\xspace}

\hypersetup{colorlinks=true,citecolor=blue,urlcolor=black,linkcolor=blue}
\begin{document}
\newcommand{\kg}{\ensuremath{\mbox{kg}}\xspace}
\newcommand{\eV}{\ensuremath{\mbox{e\kern-0.1em V}}\xspace}
\newcommand{\GeV}{\ensuremath{\mbox{Ge\kern-0.1em V}}\xspace}
\newcommand{\MeV}{\ensuremath{\mbox{Me\kern-0.1em V}}\xspace}
\newcommand{\byc}{\kern-0.1em/\kern-0.1em c}
\newcommand{\GeVc}{\ensuremath{\mbox{Ge\kern-0.1em V}\byc}\xspace}
\newcommand{\GeVcc}{\ensuremath{\mbox{Ge\kern-0.1em V}\byc^2}\xspace}
\newcommand{\MeVcc}{\ensuremath{\mbox{Me\kern-0.1em V}\byc^2}\xspace}
\newcommand{\AGeV}{\ensuremath{A\,\mbox{Ge\kern-0.1em V}}\xspace}
\newcommand{\AGeVc}{\ensuremath{A\,\mbox{Ge\kern-0.1em V}\byc}\xspace}
\newcommand{\MeVc}{\ensuremath{\mbox{Me\kern-0.1em V}\byc}\xspace}
\newcommand{\T}{\ensuremath{\mbox{T}}\xspace}
\newcommand{\cmsq}{\ensuremath{\mbox{cm}^2}\xspace}
\newcommand{\msq}{\ensuremath{\mbox{m}^2}\xspace}
\newcommand{\cm}{\ensuremath{\mbox{cm}}\xspace}
\newcommand{\mm}{\ensuremath{\mbox{mm}}\xspace}
\newcommand{\micron}{\ensuremath{\mu\mbox{m}}\xspace}
\newcommand{\mrad}{\ensuremath{\mbox{mrad}}\xspace}
\newcommand{\ns}{\ensuremath{\mbox{ns}}\xspace}
\newcommand{\m}{\ensuremath{\mbox{m}}\xspace}
\newcommand{\s}{\ensuremath{\mbox{s}}\xspace}
\newcommand{\ms}{\ensuremath{\mbox{ms}}\xspace}
\newcommand{\ps}{\ensuremath{\mbox{ps}}\xspace}
\newcommand{\dd}{\ensuremath{{\textrm d}}\xspace}
\newcommand{\dedx}{\ensuremath{\dd E\!~/~\!\dd x}\xspace}
\newcommand{\tofdedx}{\ensuremath{\textup{\emph{tof}}-\dd E/\dd x}\xspace}
\newcommand{\tof}{\ensuremath{\textup{\emph{tof}}}\xspace}
\newcommand{\pt}{\ensuremath{p_{\textrm T}}\xspace}
\newcommand{\PT}{\ensuremath{P_\textup{T}}\xspace}
\newcommand{\mt}{\ensuremath{m_{\textrm T}}\xspace}
\newcommand{\y}{\ensuremath{{y}}\xspace}

\newcommand{\pbar}{\ensuremath{\overline{\textit{p}}}\xspace}
\newcommand{\nbar}{\ensuremath{\overline{\textit{n}}}}
\newcommand{\dbar}{\ensuremath{\overline{\textup{d}}}}
\newcommand{\pim}{\ensuremath{\pi^-}\xspace}
\newcommand{\pip}{\ensuremath{\pi^+}\xspace}
\newcommand{\km}{\ensuremath{\textit{K}^-}\xspace}
\newcommand{\kp}{\ensuremath{\textit{K}^+}\xspace}
\newcommand{\hm}{\ensuremath{\textit{h}^-}\xspace}
\newcommand{\Xim}{\ensuremath{\Xi^-}\xspace}
\newcommand{\Xip}{\ensuremath{\overline{\Xi}^+}\xspace}

\newcommand{\pp}{\mbox{\textit{p+p}}\xspace}
\newcommand{\pA}{\mbox{\textit{p}+A}\xspace}
\newcommand{\NN}{\mbox{\textit{N+N}}\xspace}


\def\Offline{\mbox{$\overline{\text%
{Off}}$\hspace{.05em}\raisebox{.4ex}{\underline{line}}}\xspace}
\def\SHOE{\mbox{SHO\hspace{-1.34ex}\raisebox{0.2ex}{\color{green}\textasteriskcentered}\hspace{0.25ex}E}\xspace}
\def\DSHACK{\mbox{DS\hspace{0.15ex}$\hbar$ACK}\xspace}
\DeclareRobustCommand{\SHINE}{\mbox{\textsc{S\hspace{.05em}\raisebox{.4ex}{\underline{hine}}}}\xspace} 
\def\Glissando{\textsc{Glissando}\xspace}
\newcommand{\FlukaLong}{{\scshape Fluka2008}\xspace}
\newcommand{\FlukaEleven}{{\scshape Fluka2011}\xspace}
\newcommand{\Fluka}{{\scshape Fluka}\xspace}
\newcommand{\UrqmdLong}{{\scshape U}r{\scshape qmd1.3.1}\xspace}
\newcommand{\Urqmd}{{\scshape U}r{\scshape qmd}\xspace}
\newcommand{\GheishaLong}{{\scshape Gheisha2002}\xspace}
\newcommand{\GheishaOld}{{\scshape Gheisha600}\xspace}
\newcommand{\Gheisha}{{\scshape Gheisha}\xspace}
\newcommand{\Corsika}{{\scshape Corsika}\xspace}
\newcommand{\Venus}{{\scshape Venus}\xspace}
\newcommand{\VenusLong}{{\scshape Venus4.12}\xspace}
\newcommand{\GiBUU}{{\scshape GiBUU}\xspace}
\newcommand{\GiBUULong}{{\scshape GiBUU1.6}\xspace}
\newcommand{\FlukaNewLong}{{\scshape Fluka2011.2\_17}\xspace}
\newcommand{\Root}{{\scshape Root}\xspace}
\newcommand{\Geant}{{\scshape Geant}\xspace}
\newcommand{\GeantThree}{{\scshape Geant3}\xspace}
\newcommand{\GeantFour}{{\scshape Geant4}\xspace}
\newcommand{\QGSJet}{{\scshape QGSJet}\xspace}
\newcommand{\DPMJet}{{\scshape DPMJet}\xspace}
\newcommand{\Epos}{{\scshape Epos}\xspace}
\newcommand{\EposLong}{{\scshape Epos1.99}\xspace}
\newcommand{\QGSJetLong}{{\scshape QGSJetII-04}\xspace}
\newcommand{\DPMJetLong}{{\scshape DPMJet3.06}\xspace}
\newcommand{\SibyllLong}{{\scshape Sibyll2.1}\xspace}
\newcommand{\EposLHCLong}{{\scshape EposLHC}\xspace}
\newcommand{\Hsd}{{\scshape Hsd}\xspace}
\newcommand{\Ampt}{{\scshape Ampt}\xspace}
\newcommand{\Hijing}{{\scshape Hijing}\xspace}
\newcommand{\PHSD}{{\scshape Phsd}\xspace}
\newcommand{\SmashModel}{{\scshape Smash}\xspace}

\def\red#1{{\color{red}#1}}
\def\blue#1{{\color{blue}#1}}
\def\avg#1{\langle{#1}\rangle}
\def\sci#1#2{#1\!\times\!10^{#2}}
\newcommand{\Fi}[1]{Fig.~\ref{#1}}
\newcommand{\CernVM}{\textsc{Cern\-\kern-0.05emVM}\xspace}

\def \pions{$\pi^\pm$\xspace}
\def \kaons{K$^\pm$\xspace}
\def \proton{p\xspace}
\def \antiproton{$\bar{\text{p}}$\xspace}
\def \protons{p($\bar{\text{p}}$)\xspace}
\def \lamb{$\Lambda$\xspace}
\def \antilamb{$\bar{\Lambda}$\xspace}
\def \lambs{$\Lambda(\bar{\Lambda})$\xspace}
\def \kzeros{K$_{S}^{0}$\xspace}
\def \ncl{$N_{\mathrm{cl}}$\xspace}
\def \shine{NA61/SHINE}
\def \pipi{$\pi^+\pi^-$\xspace}
\def \vzero{$V^0$\xspace}
\def \kaonstar{K$^{*0}$\xspace}
\def \rhozero{$\rho^{0}$\xspace}
\newcommand{\pT}{\ensuremath{p_\text{T}}\xspace}
\newcommand{\TeV}{\ensuremath{\mbox{Te\kern-0.1em V}}\xspace}
\newcommand{\snn}{\ensuremath{\sqrt{s_\mathrm{NN}}}\xspace}
\title{News on strangeness production from the \shine \\ experiment}
%
%

\author{\firstname{Yuliia} \lastname{Balkova}\inst{1}\fnsep\thanks{\email{yuliia.balkova@cern.ch}} \and
        \firstname{Tatjana} \lastname{Šuša}\inst{2}\fnsep\thanks{\email{tatjana.susa@cern.ch}}
        for the \shine~Collaboration
}

\institute{National Centre for Nuclear Research, Pasteura 7 str., 02-093 Warsaw, Poland
\and
Ruđer Bošković Institute, Bijenička cesta 54, 10000 Zagreb, Croatia
}

\abstract{Strangeness production in high-energy hadronic and nuclear collisions continues to be one of the central topics in the study of strongly interacting matter. The data collected by the NA61/SHINE experiment at the CERN SPS North Area allow for a comprehensive scan of the strangeness production across various collision energies and system sizes.

This article focuses on the new results of the strangeness production in central collisions of medium-sized nuclei, such as Ar+Sc, at the SPS energy range. In particular, the results for Lambda hyperons and charged and neutral $K$ mesons are shown. The energy and system size dependencies of Lambda-to-pion and strangeness-to-pion ratios are also explored. Moreover, an unexpected excess of charged over neutral $K$ meson production in Ar+Sc and $\pi^-$+C interactions is presented. The obtained results are compared with predictions from selected particle production models, as well as with existing world data from proton-proton and nucleus-nucleus collisions.}
\maketitle
\vspace{-0.3cm}
\section{Introduction}
\label{intro}

The study of strongly interacting matter under extreme conditions provides valuable insights into the nature of the early Universe and the structure of compact astrophysical objects. In particular, high-energy heavy-ion collisions allow experimental access to the transition between hadronic matter and the quark-gluon plasma (QGP) and, thus, outline the phase diagram of strongly interacting matter.

One of the earliest proposed signatures of QGP formation is the enhancement of strangeness production relative to baseline proton-proton and proton-nucleus collisions~\cite{Rafelski:1982pu}. Various experiments at the CERN Super Proton Synchrotron (SPS), including NA49, reported striking evidence of enhanced strangeness in heavy-ion collisions. In particular, NA49 experiment observed non-monotonic behaviour in the energy dependence of the $K^+/\pi^+$ at intermediate SPS energies, collectively referred to as the \textit{``horn''} structure~\cite{Alt:2007aa}.

Nevertheless, data for intermediate-size systems remains insufficient. Such studies serve as valuable tests for theoretical models of particle production and are crucial for disentangling collective effects associated with QGP formation from initial-state effects present in smaller systems. The NA61/SHINE experiment at the CERN SPS addresses this gap by systematically studying collisions across a wide range of system sizes and beam energies. It employs a fixed-target setup with large acceptance and excellent tracking and particle identification capabilities~\cite{Abgrall:2014xwa}.

\section{Results}
\label{results}

\subsection{\pmb{$\Lambda$} baryon production in Ar+Sc collisions}
New preliminary results on $\Lambda$ baryon production in the 10\% most central Ar+Sc collisions at beam momenta of 19$A$ and 30\AGeVc (corresponding to collision center-of-mass energy per nucleon pair \snn = 6.12, 7.62~GeV) are presented along with earlier results at three higher energies (40$A$, 75$A$, and 150\AGeVc, which correspond to \snn = 8.77, 11.9, and 16.8 GeV, respectively) previously shown in Ref.~\cite{Balkova:2024dje}. $\Lambda$ baryons were reconstructed via their dominant weak decay channel $\Lambda \rightarrow p + \pi^-$ with branching ratio BR = 63.9\%. 
The results discussed below refer to hadrons produced in strong interactions and electromagnetic decays. The raw yields, obtained by fitting the invariant mass spectra, are corrected for detector acceptance, reconstruction efficiency, selection cuts, branching ratio, and feed-down from heavier hyperons. For full details of the analysis procedure, see Ref.~\cite{Balkova:2022fps}. The obtained mean multiplicities of $\Lambda$ baryons are 4.073 $\pm$ 0.052 (stat) $\pm$ 0.24 (sys) for Ar+Sc at 19\AGeVc, and 5.190 $\pm$ 0.059 (stat) $\pm$ 0.36 (sys) for Ar+Sc at 30\AGeVc.

\begin{figure}[ht]
  \centering
  \includegraphics[width=5cm,clip]{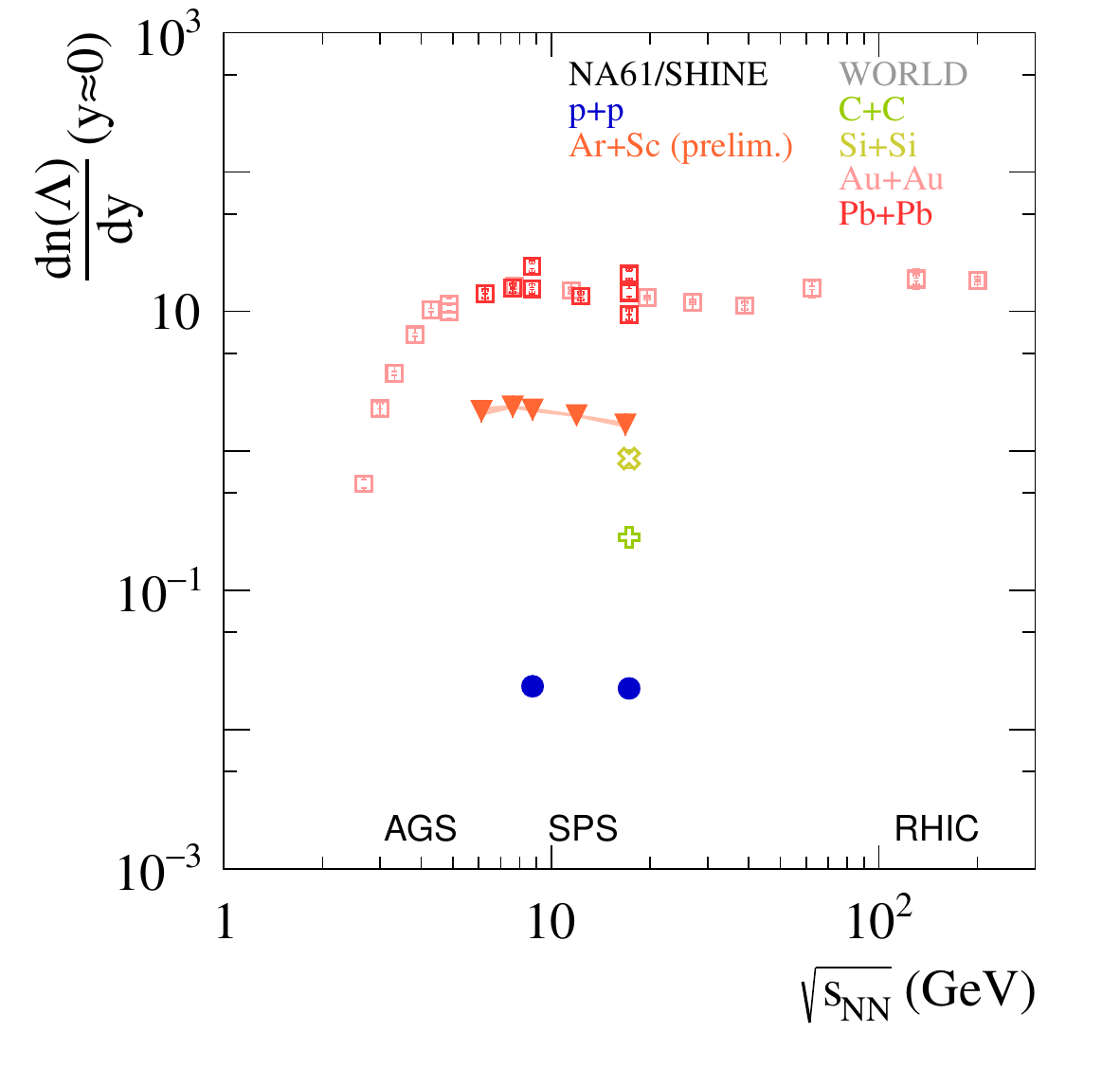}
  \includegraphics[width=5cm,clip]{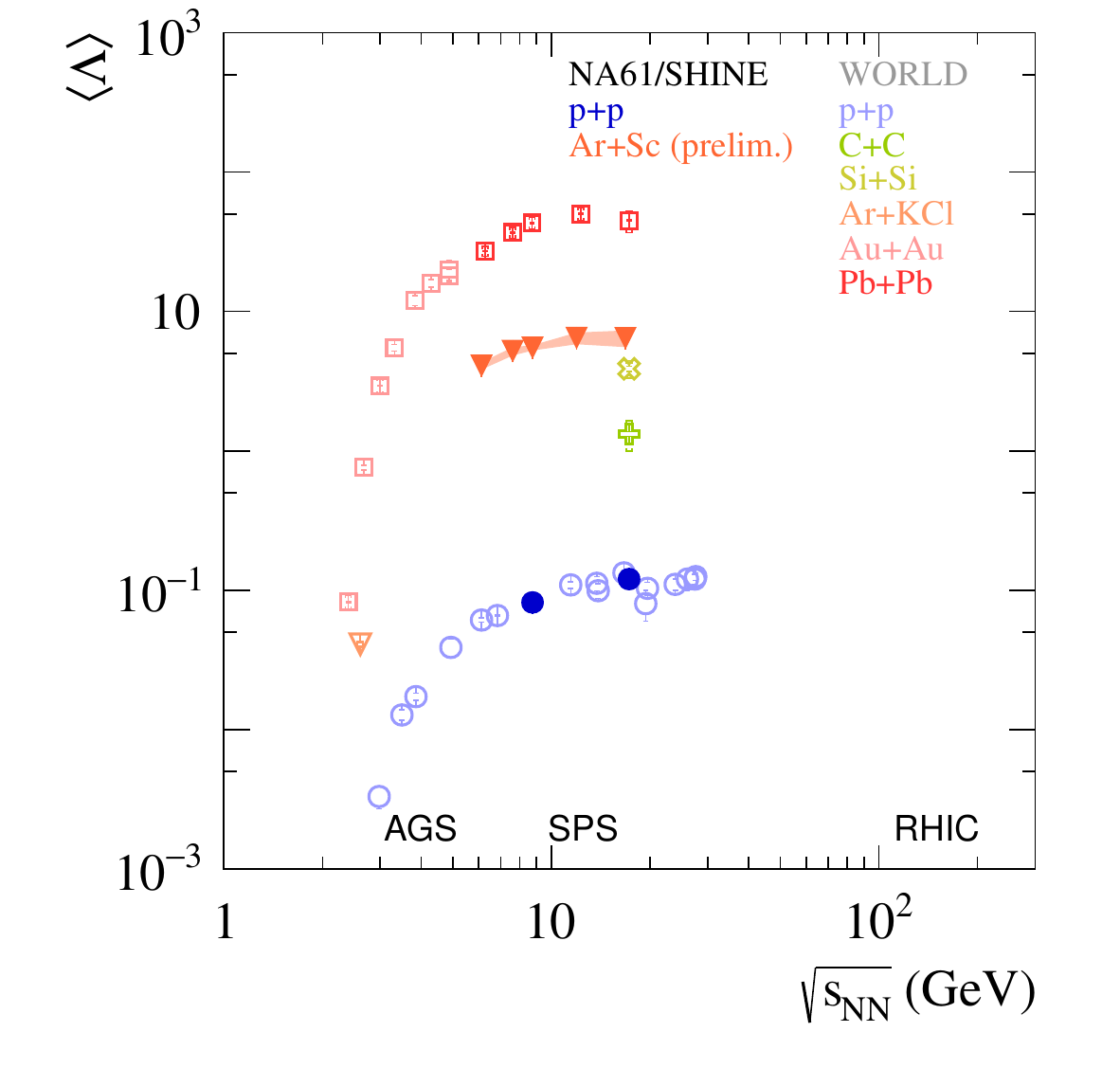}
  \caption{The energy dependence of mid-rapidity yield (\textit{left}) and the mean multiplicity (\textit{right}) of $\Lambda$ baryons. The systematic uncertainties of the Ar+Sc results are presented as shaded bands. Results for \pp, Ar+Sc, Ar+KCl, C+C, Si+Si, Au+Au, and Pb+Pb are shown (NA61/SHINE: \pp, Ar+Sc; NA49: C+C, Si+Si, Pb+Pb; NA57: Pb+Pb; STAR: Au+Au; PHENIX: Au+Au; E891: Au+Au; E895: Au+Au; E896: Au+Au; HADES: Ar+KCl, Au+Au; bubble chamber experiments: \pp). For the references to the world data, see Ref.~\cite{QM_pres}.}
  \label{fig:lambda}
\end{figure}

Figure~\ref{fig:lambda} illustrates the energy dependence of the mid-rapidity yields and mean multiplicities of $\Lambda$ baryons in Ar+Sc collisions, compared against existing measurements from a range of systems. Within the SPS energy range, both observables tend to flatten, indicating a saturation-like behaviour that appears to be independent of system size. Interestingly, the values obtained in Ar+Sc and Si+Si collisions are much closer to those measured in central Au+Au and Pb+Pb interactions than to those observed in \pp collisions.

\begin{figure}[ht]
  \centering
  \includegraphics[width=5cm,clip]{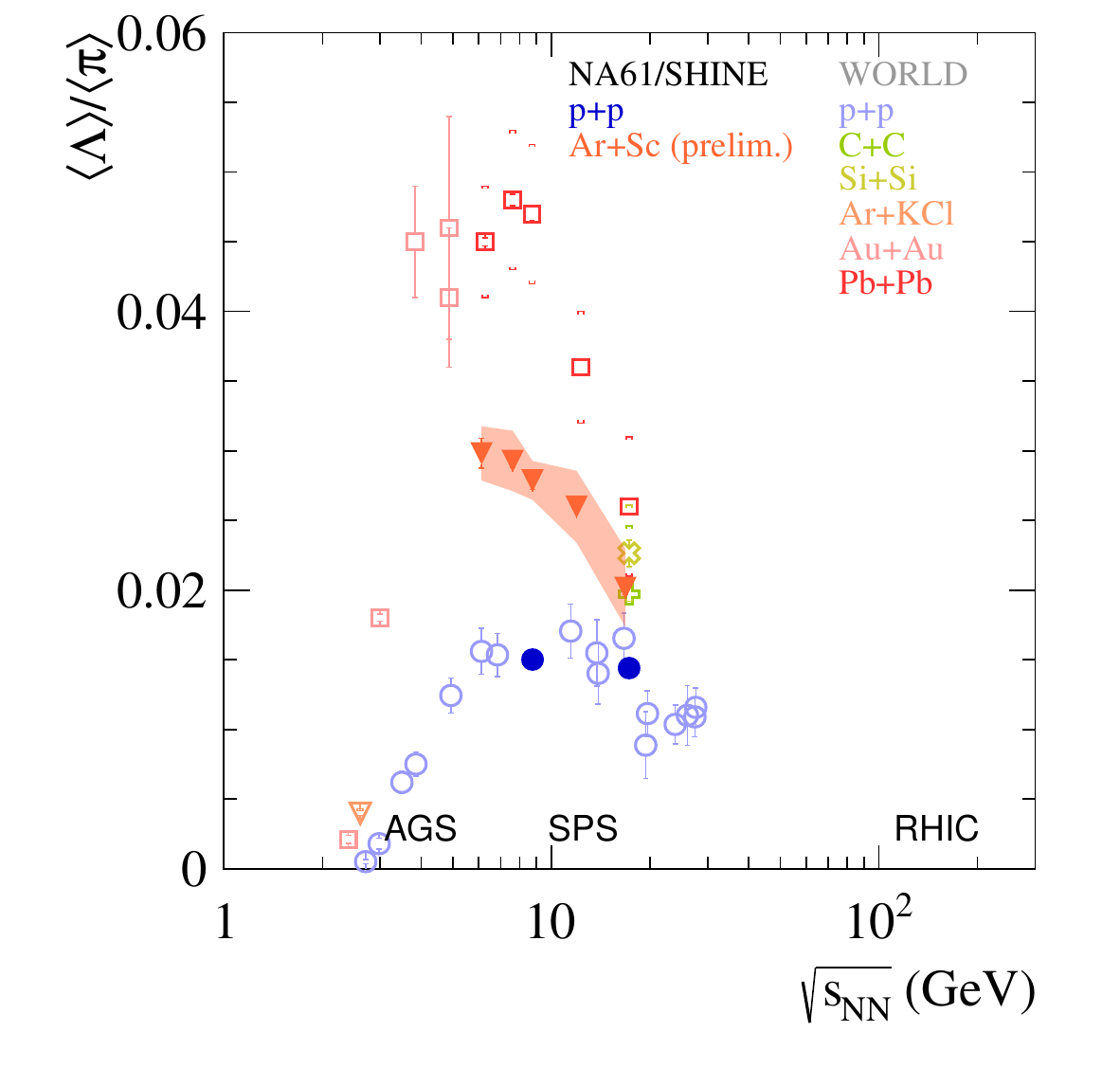}
  \includegraphics[width=5cm,clip]{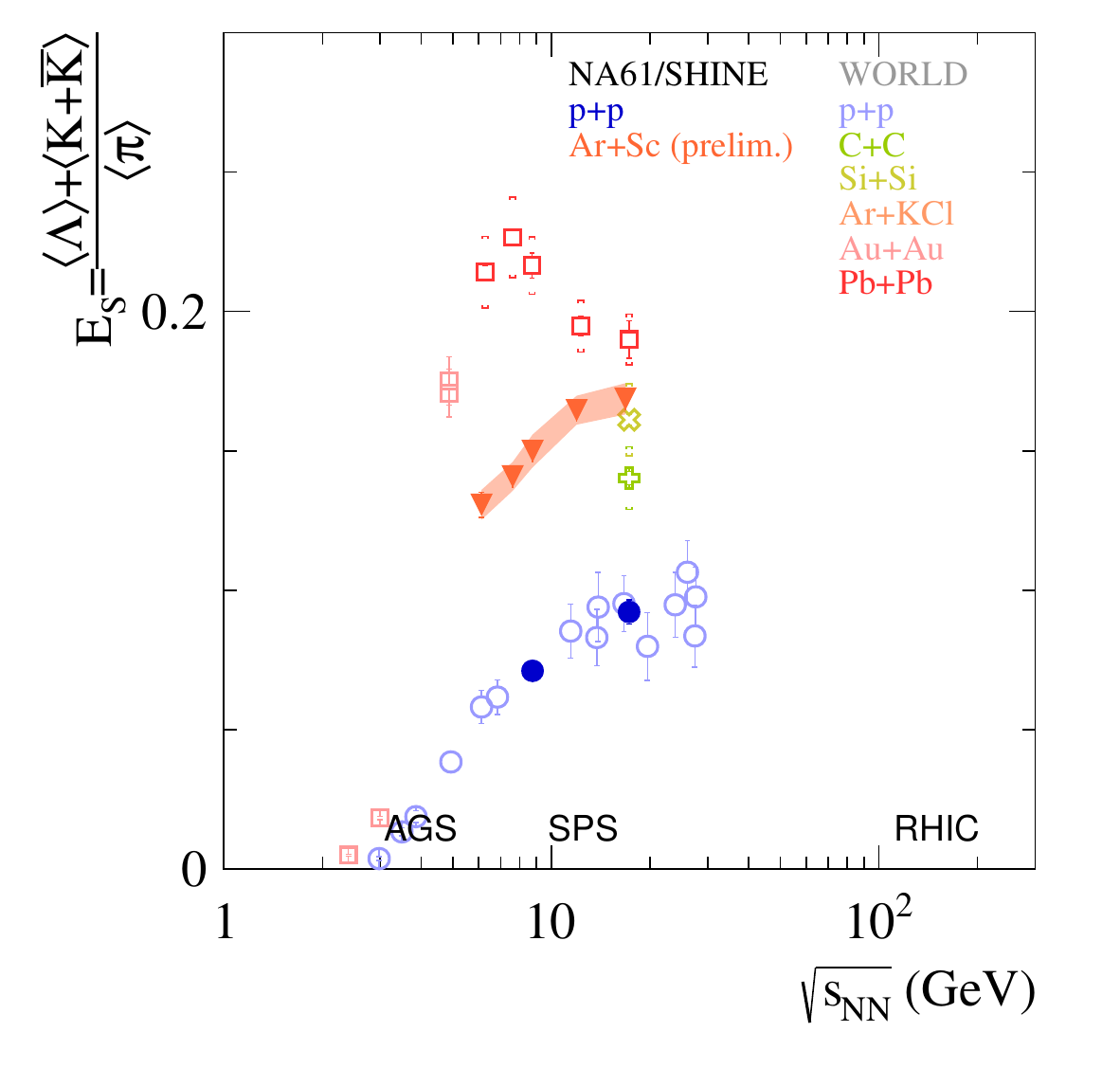}
  \caption{The energy dependence of $\langle \Lambda \rangle / \langle \pi \rangle$ ratio (\textit{left}) and strangeness enhancement $E_S$ (\textit{right}). The systematic uncertainties of the Ar+Sc results are presented as shaded bands. Results for \pp, Ar+Sc, Ar+KCl, C+C, Si+Si, Au+Au, and Pb+Pb are shown (NA61/SHINE: \pp, Ar+Sc; NA49: C+C, Si+Si, Pb+Pb; NA57: Pb+Pb; STAR: Au+Au; PHENIX: Au+Au; E891: Au+Au; E895: Au+Au; E896: Au+Au; HADES: Ar+KCl, Au+Au; bubble chamber experiments: \pp). For the references to the world data, see Ref.~\cite{QM_pres}.}
  \label{fig:strangeness}
\end{figure}

Figure~\ref{fig:strangeness} shows the energy dependence of the $\langle\Lambda\rangle/\langle\pi\rangle$ ratio and the strangeness enhancement factor $E_S$, defined following Ref.~\cite{NA35:1994veo} as $E_S = \frac{\langle \Lambda \rangle + \langle K + \overline{K} \rangle}{\langle \pi \rangle}$, where $\langle \pi \rangle$ is equal to $1.5\cdot(\langle \pi^+ \rangle + \langle \pi^- \rangle)$, and $\langle\pi^+\rangle$ and $\langle\pi^-\rangle$ are mean multiplicities of positively and negatively charged pions, respectively. Since the $\overline{\Lambda}/\Lambda$ ratio is below 0.15 in the SPS energy range~\cite{Alt:2007aa}, the contribution of $\overline{\Lambda}$ to $E_S$ is neglected. For \pp interactions, $\langle K + \overline{K} \rangle$ is approximated as $4 \cdot \langle K^0_S \rangle$, while for nucleus-nucleus collisions it is calculated as $2 \cdot (\langle K^+ \rangle + \langle K^- \rangle)$ as suggested in Ref.~\cite{Gazdzicki:1996pk}. In central Pb+Pb and Au+Au collisions, both $\langle\Lambda\rangle/\langle\pi\rangle$ and $E_S$ display a pronounced maximum near mid-SPS energies. No such peak is observed for lighter systems, including Ar+Sc, though the overall trend in $\langle\Lambda\rangle/\langle\pi\rangle$ in Ar+Sc is comparable to that in Pb+Pb collisions.

\subsection{Charged and neutral kaon productions in Ar+Sc collisions}
NA61/SHINE has measured charged kaon production in the 10\% most central Ar+Sc collisions at beam momenta of 13$A$, 19$A$, 30$A$, 40$A$, 75$A$, and 150\AGeVc (corresponding to \snn = 5.12, 6.12, 7.62, 8.77, 11.9 and 16.8 GeV)~\cite{ChargedKaonRef}. Charged kaons were identified by using \mydedx method and combined \tof-\mydedx method. The yields of charged $K$ mesons  at mid-rapidity in the 10\% most central Ar+Sc collisions at beam momenta of 40$A$ and 75\AGeVc were determined in the interval of collision center-of-mass rapidity 0.0 $< y < $ 0.2 and amount to $(\mathrm{d}n/\mathrm{d}y)(\kp)_{y \approx 0} = $ 3.732 $\pm$ 0.016 (stat) $\pm$ 0.148 (sys) and $(\mathrm{d}n/\mathrm{d}y)(\km)_{y \approx 0} = $ 2.029 $\pm$ 0.012 (stat) $\pm$ 0.069 (sys) at 75\AGeVc and $(\mathrm{d}n/\mathrm{d}y)(\kp)_{y \approx 0} = $ 3.283 $\pm$ 0.041 (stat) $\pm$ 0.118 (sys) and $(\mathrm{d}n/\mathrm{d}y)(\km)_{y \approx 0} = $ 1.417 $\pm$ 0.024 (stat) $\pm$ 0.048 (sys) at 40\AGeVc.

\begin{figure}[ht]
\centering
\includegraphics[width=5cm,clip]{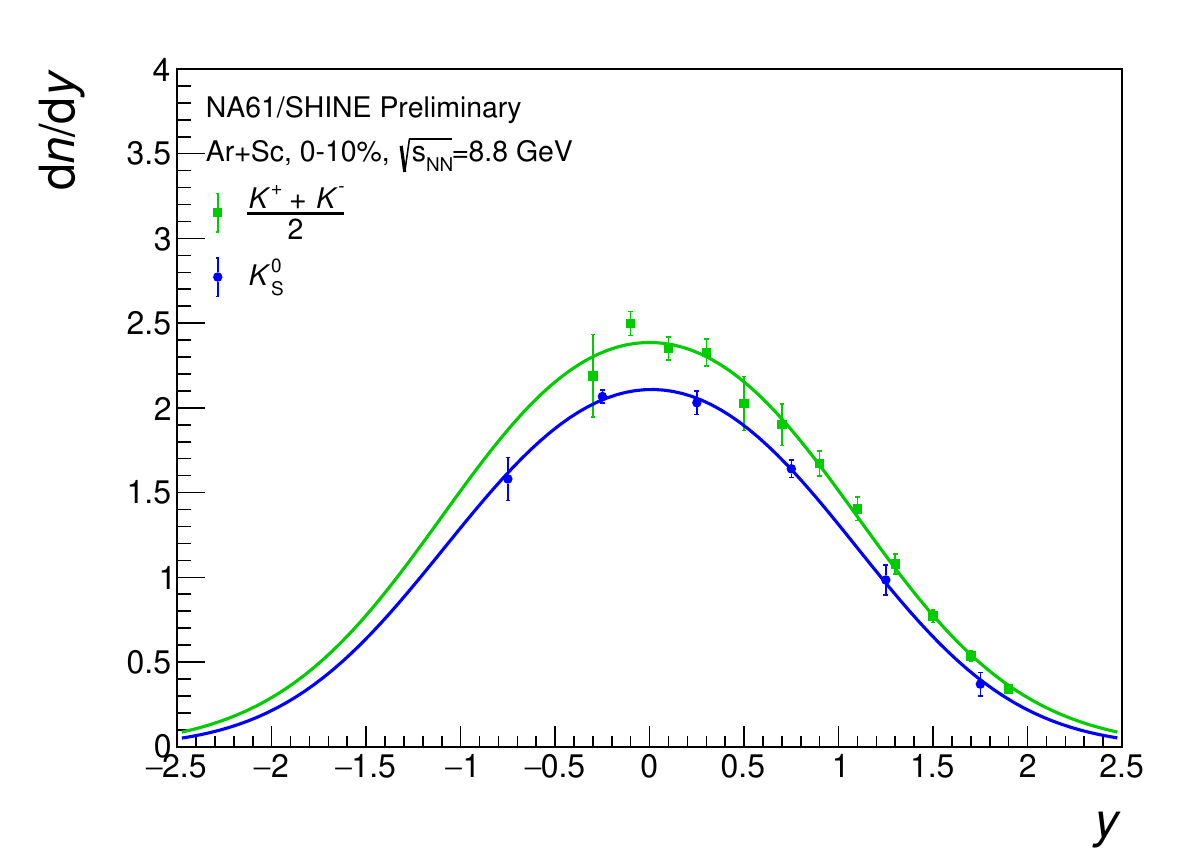}
\includegraphics[width=7cm,clip]{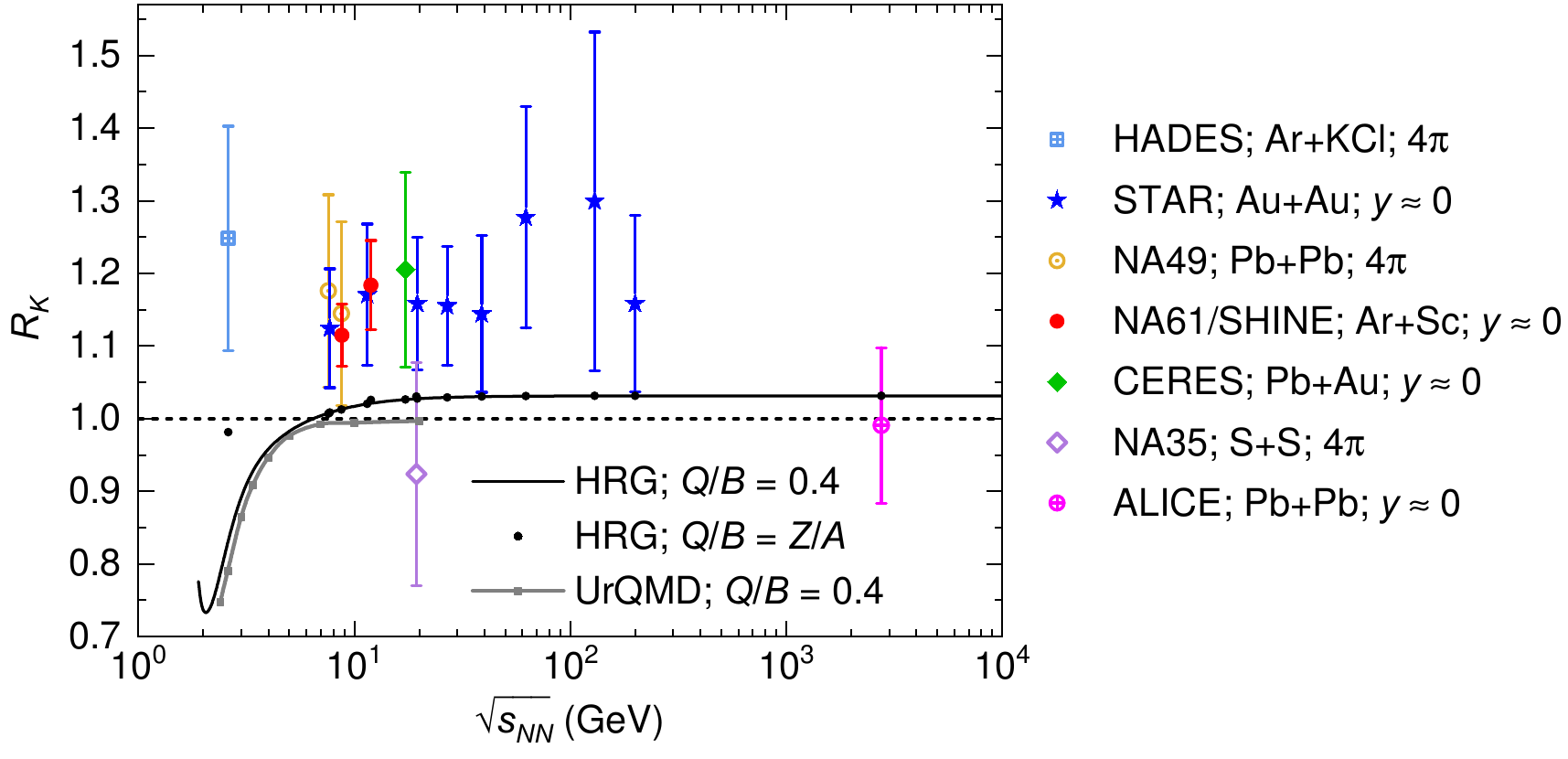}
\caption{
\textit{Left}: The rapidity spectrum of \mykzeros mesons (blue circles, preliminary) and the averaged spectrum of charged kaons (green squares, Ref.~\cite{ChargedKaonRef}) measured in 0--10\% central Ar+Sc collisions at 40\AGeVc. Total uncertainties are drawn. Blue and green curves show the result of a double Gaussian fit to the \mykzeros and averaged fit for charged kaons spectra, respectively. \textit{Right}: Charged-to-neutral kaon ratio as a function of collision energy. Experimental data are shown by symbols with total uncertainties. HRG baseline for electric-to-baryon charge $Q/B$ = 0.4 is shown by a black line. HRG baseline for $Q/B$ values specified accordingly to the given types of colliding nuclei is represented by black dots. UrQMD model results are shown by grey squares. Figure from Ref.~\cite{NA61SHINE:2023azp} updated with the new NA61/SHINE data point.}
\label{K0s}
\end{figure}

\mykzeros mesons were identified by the characteristic topology of their decay into charged pions (\mykzeros $ \rightarrow \pi^{+} + \pi^{-}$) with a branching ratio BR = 69.2\%. Their production was measured in the 10\% most central Ar+Sc collisions at beam momenta of 40$A$ (preliminary) and 75\AGeVc~\cite{NA61SHINE:2023azp}. Figure~\ref{K0s} (\textit{left}) shows the rapidity distribution of \mykzeros mesons at beam momentum of 40\AGeVc, together with the result of a double Gaussian fit. The yield of \mykzeros meson at mid-rapidity,  determined from the fit at $y =$ 0,  amounts to $(\mathrm{d}n/\mathrm{d}y)_{y \approx 0} = $ 2.108 $\pm$ 0.016 (stat) $\pm$ 0.052 (sys) and $(\mathrm{d}n/\mathrm{d}y)_{y \approx 0} = $ 2.433 $\pm$ 0.027 (stat) $\pm$ 0.102 (sys) at 40$A$ and 75\AGeVc, respectively. At both momenta, an excess of charged over neutral kaons is observed, although an approximately equal abundance of charged and neutral kaons is expected since Ar and Sc nuclei are nearly isospin symmetric (valence quarks u = d within 6\%), for details see Ref.~\cite{NA61SHINE:2023azp}. At mid-rapidity, the charged-to-neutral kaon ratio, defined as $R_{K} = (\kp + \km)/(2K^{0}_{S})$, was found to be 1.115 $\pm$ 0.043 (tot) and 1.184 $\pm$ 0.061 (tot) at 40$A$ and 75\AGeVc, respectively.

In Fig.~\ref{K0s} (\textit{right}), the charged-to-neutral kaon ratio is compared with a compilation of kaon production results from other experiments and predictions from the Ultrarelativistic Quantum Molecular Dynamics (UrQMD)~\cite{UrQMD_Ref} and Hadron Resonance Gas (HRG)~\cite{HRG_Ref} models. While data from other experiments are consistent with the \shine~results, both models fail to reproduce the observed charged-to-neutral kaon ratio. The deviation of the experimental results for the $R_K$ factor from HRG model prediction reaches a significance of 4.7$\sigma$ when only the NA61/SHINE data at 75\AGeVc are included in the averaging (see Ref.~\cite{NA61SHINE:2023azp} for details). This significance increases to 5.3$\sigma$ when the NA61/SHINE results at 40\AGeVc are also taken into account. 

Ratios $R_{K}$ > 1  were also observed in $\pi^{-}$+C collisions at beam momenta of 158 and 350 \GeVc~\cite{piCRef}. The models fail to reproduce the charged-to-neutral kaon ratio also for this small asymmetric system.

%

\end{document}